\documentstyle[aps,prl,psfig,preprint]{revtex}
              
\begin{document}
\bibliographystyle{prsty}
\title{Damage-Spreading in Self-Organized Critical Systems}
\author{Raffaele Cafiero$^a$ Angelo Valleriani$^b$ and Jos\'e Luis Vega$^c$}
\address{$^a$P.M.M.H. Ecole Sup\'erieure de Physique et de 
Chimie Industrielles, 10 rue Vaquelin, 75231 Paris, France\\
$^b$Max-Planck-Institut f\"ur Kolloid- und Grenzfl\"achenforschung, 
D-14424 Potsdam, Germany\\$^c$Max-Planck-Institut f\"ur Physik komplexer Systeme,
N\"othnitzer Str.\ 38, D-01187 Dresden, Germany}
\maketitle

\begin{abstract}
\noindent
We study the behavior under perturbations of different versions of  
Bak-Sneppen (BS) model  in $1\! +\! 1$ dimension. We  focus our 
attention on 
the damage-spreading features 
of the BS  model  with both random as well as deterministic updating,
with sequential as well as parallel updating.  
 In addition, we compute analytically the 
asymptotic plateau reached by the distance after the growing phase.
\end{abstract}
\vspace{1cm}
\vfill

\section{Introduction}
A great deal of evidence has been put forward 
in recent years for the appearance of
 power law statistics in nature: A wide variety of 
phenomena, from earthquakes \cite{CJ89,GS96} to
biological evolution \cite{BS93,Sole,MV98,AM99,Drossel98}, 
from 
surface growth \cite{Sneppen92}
to fluid displacement in porous media \cite{WW83CR88},
exhibit  scale invariance in both space and time.
 To explain these all-pervading  power-law tails, Bak, Tang, 
and Wiesenfeld introduced  the 
 concept of self-organized criticality (SOC) \cite{BTW87}. 
In a nutshell, SOC means that
certain driven spatially extended systems evolve 
spontaneously towards a
critical globally stationary dynamical state 
with no characteristic time or length scales \cite{SJD95}.
This scale invariance implies that the 
correlation length in these systems
is infinite and consequently a small (local)
 perturbation
can produce a global (maybe even drastic) effect.
This possibility leads naturally to the study 
of the sensitivity to perturbations 
in  (self-organized) critical systems.

To study the propagation of local 
perturbations ({\em damage spreading})
one can borrow a technique 
from dynamical systems theory. Let us consider, 
for instance, two copies of the same dynamical system 
(let us say, for instance, a 1D map) , 
 with slightly different initial
conditions.  By following the dynamics of  both copies  and 
studying the evolution in time of 
the ``distance'' $d(t)$  between them, it is
possible to 
quantify  the effect of the initial perturbation.
Indeed, assuming that the distance $d(t)$ grows exponentially, 
and defining the 
Lyapunov exponent $\lambda$  via  
\begin{equation}
\label{a}
D(t) = D_0 \exp(\lambda\, t),
\end{equation}
three different 
behaviors can be distinguished,  corresponding to  $\lambda$ being 
either positive, negative or zero. 
The case $\lambda>0$ corresponds to the so-called 
{\em chaotic} systems, where the extremely high 
sensibility to initial conditions leads to
exponentially diverging trajectories on a {\em strange} (or chaotic) 
attractor. The case $\lambda<0$, 
instead,  characterizes
 those systems in which the dynamics 
has an attractor and any initial 
perturbation is ``washed out'' with exponential rapidity. 

The boundary case, $\lambda=0$,  admits, in turn, a whole class of 
functions $D(t)$, namely 
\begin{equation}
\label{a1}
D(t)\, \sim\, t^{\alpha}\, .
\end{equation}
where $\alpha$ is some exponent, characteristic of the system. 
In particular, $\alpha > 0$ corresponds 
to weak sensitivity to initial
conditions while  $\alpha < 0$ 
corresponds to weak insensitivity to initial
conditions (as an example,
the reader is referred to Refs. \cite{TPZ97}, where this analysis 
is performed for the logistic map at its critical 
point \cite{Note0}). 

Recently \cite{TCT98,VV98,CVV98,CVV99}, a similar analysis was
performed on several versions of the Bak-Sneppen (BS) model \cite{BS93}.
Originally proposed  to describe  ecological evolution,
this model has been paid a great deal of attention
due to its simplicity
and the fact that it exhibits self-organized criticality.

Schematically, the BS model is defined on a lattice
 where at each time-step one site is chosen, namely the one
that fulfills a global constraint
(minimum in
some phase space). This site is defined as the {\it active site}.
This
dynamics leads to a non-Markovian process where the activity, i.e.\ the
position on the lattice of the active site, jumps on the lattice following a
(correlated) L\'evy Walk.
Its critical properties  allow us to describe its behavior under perturbations
via Eq.\ (\ref{a1}),  with
\begin{equation}
\alpha\, =\, 0.32\, .
\end{equation}

In this paper we analyze in detail the behavior of different versions of the
BS model.  As we shall see, the particular stationary
distribution of the variable does not play any role in the determination of
the exponent $\alpha$ in Eq.\ (\ref{a1}). What matters is, instead, the kind of
L\'evy Walk involved and the strength of the correlations.

The paper is organized as follows.  After a general 
description of our formalism in Sec.\
\ref{general},
the BS model is discussed in Sec.\ \ref{bsmodel}. 
In Sec.\ \ref{deterministic} we analyze  
the behavior under perturbations of 
the deterministic BS model, introduced  
in \cite{DVV97,DVV98}. 
In Sec. \ref{parallel}, we study the parallel version of it (PBS model),
 introduced not long ago by Sornette 
and Dornic \cite{SD96}. We  study  two versions of it, namely
the one with random updating and the one with
logistic updating. 
We also present here a discussion of the differences and similarities 
between the BS and the PBS models. In Sec.\ \ref{pert} we discuss the
influence of changing the perturbation. 
Conclusions can be found in Sec.\ \ref{conclusions}.

\section{General formalism: Damage-Spreading in the Ring}
\label{general}
To illustrate our formalism, let us start by considering a lattice 
of $N$ sites on  a one-dimensional ring $R_1$. To each  site $j$ 
we assign  a random number $f_j$, extracted from a  uniform
distribution between $0$ and $1$. 
We then consider a ``replica'' $R_2$, in which  we introduce a 
perturbation by  exchanging  the positions of 
the values of $f_{k_1}$  and $f_{k_2}$.
We define as active the sites $k_1$ in $R_1$ and (the randomly chosen site) 
$k_2$ in $R_2$, namely those sites at which we have the same value of $f$.
It is clear that, from a statistical point of view,  both system $R_1$ and 
replica $R_2$ are described by the same distribution function. 
The prescription just introduced corresponds, in a suitably 
defined phase space, to a small difference in the initial conditions 
between $R_1$ and $R_2$ (see Eq.\ (\ref{c11}) below).
Moreover,  this procedure of finding an active site and 
exchanging its position with another site taken at 
random along the lattice,  corresponds exactly to the one 
proposed in \cite{TCT98} for the Bak-Sneppen model (other definitions for 
the initial perturbation are considered elsewhere \cite{CVV98}). 
The dynamics on the ring(s) is defined as follows. At each time-step,
an integer random number $x$ between $1$ and $N$ is chosen. 
Bearing  in mind that our rings have
periodic boundary conditions, the positions of the new active sites at time
$t+1$ is then given by 
\begin{eqnarray}
\label{a2b}
k_1(t+1) & = & k_1(t)+x \\ 
k_2(t+1) & = & k_2(t)+x
\label{a2} 
\end{eqnarray}
on the rings $R_1$ and $R_2$ respectively. On these active  sites, 
the value of the variables $f$ is changed, assigning to both  
of them the same random  number (this corresponds to the choice of  the same
sequence of random numbers in \cite{TCT98} or 
to the same thermal noise in usual damage 
spreading calculations \cite{SSKH87}). 

As both system $R_1$ and replica $R_2$ evolve, we compute 
 the Hamming distance, namely
\begin{equation}
\label{b}
D(t)\, =\, \frac{1}{N}\sum_{j=1}^N\, |f_j^1-f_j^2|\, .
\end{equation}
Since this quantity has strong fluctuations, 
we  will consider  the average 
$\langle D(t)\rangle$, over realizations. In particular, at $t=1$,
the average (initial) distance $\langle D(1)\rangle$ can be 
obtained from Eq.\ (\ref{b}), 
\begin{equation}
\label{b22}
\langle D(1)\rangle \, = \, \frac{2}{N}
\int_0^1 \, df_1 df_2  \, \eta_1 ( f^1 ) 
\eta_2 ( f^2 ) \,  | f^1 - f^2 | ,
\end{equation}
where $\eta_i$ is the distribution function (at $t=1$)
for the variable $f^i\in R_i$. In this toy model, both distributions 
$\eta_i(f)$ are the same, namely a uniform
distribution in  $f^i\in R_i$. A simple computation yields
\begin{equation} 
\label{c11}
\langle D(1)\rangle =\frac{2}{3N}\, . 
\end{equation}
Applying a similar procedure,  one can verify  that for
 $1\! \ll\! t\! \ll\! N$ the distance grows linearly. 
Indeed, let us define $\sigma(t)$ as the averaged 
number of different sites covered in one copy of the system
at time $1\! \ll\! t\! \ll\! N$. Then, at time $t$ only $\sigma(t)$ sites 
have been
changed and these are the only ones that contribute to distance. From this
consideration it follows that
\begin{equation}
\label{d}
\langle D(t)\rangle\,=\, \langle D(1)\rangle\,\sigma(t)\, ,
\end{equation}
where the fact that both replica
contribute to the distance on the same footing is taken into account in Eq.\
(\ref{c11}).
In the case of the ring, if $N\! \gg\! 1$ and $1\! \ll\!
 t\!\ll\! N$
the system will always choose a new site at each time-step, 
and therefore $\sigma(t)\sim
t$ (note that, in the 1D case for this dynamics, 
this is the fastest possible growth of $\langle D(t)\rangle$).
This behavior stops at times $t \sim N$ where a 
crossover to a saturation regime appears. Clearly, after $t\propto N$ time
steps 
each site of the lattice has been covered at least
once. For $t\gg N$, almost all the lattice sites have been 
covered and the two
strings are made of the same random 
numbers shifted by $k_2-k_1$. Thus, the
distance reaches a plateau,
independent on the size $N$ of the system, given by
\begin{equation}
\label{b1}
\langle D(t\to \infty)\rangle\, =\, \int_{0}^{1} df^1 df^2\,
\rho_1(f^1)\rho_2(f^2)\,|f^1-f^2|\, ,
\end{equation}
where $\rho_i$ is the normalized distribution 
function (at $t=\infty$)
for the variable $f^i\in R_i$. In Eq.\ (\ref{b1}), 
for the particular case of the ring the distributions in the integral are
given by $\rho_i=\eta_i$.
Applying Eq.\ (\ref{b1}) 
to $R_1$ and $R_2$ we finally obtain
\begin{equation}
\label{e}
\lim_{t\to\infty}\,\langle D(t)\rangle\, =\, \frac{1}{3}\, .
\end{equation}
Note that the same result can be obtained from Eqs.\ (\ref{c11},\ref{d}) once
$\sigma=N$ is inserted. 
To have an initial distance independent
of the lattice size, we consider 
the ratio $\langle D(t)\rangle/ \langle D(1)\rangle$. For this ratio, 
however, the value of the plateau 
depends linearly on $N$, i.e.\
\begin{equation}
\label{ratio}
\frac{\langle D(\infty)\rangle}{\langle D(1)\rangle} = \frac{N}{2}.
\end{equation}

In Fig.\ \ref{fig1} we show  the evolution in time of the 
ratio $\frac{\langle D(t)\rangle}{\langle D(1)\rangle}$, 
averaged over many realizations, for
 different lattice sizes. The plateau reached 
for $t\to \infty$ depends  on $N$ according to  Eq.\ (\ref{ratio}). 
The exponent $\alpha=1$ obtained for this case from Eq.\ (\ref{d}) can also be
numerically obtained with great accuracy.
From  a physical point of view, this power law behavior originates 
 in the ability of the system to cover the lattice. 
As we have seen,  the activity can jump anywhere on the lattice with 
probability $1/N$. Thus the number of sites $j$ 
with the same $f_j$ decreases linearly with time and $\langle D(t)\rangle$
increases linearly with time. As a consequence, the time $\tau$
needed to reach the plateau scales with the lattice size as $\tau\sim N$.

Keeping in mind our goal of modeling  the behavior 
of the BS model, let us now consider the case of L\'evy-Walk type activity
jumps along the lattice. More precisely, the length $x$  of any 
jump is extracted from a power law distribution function, namely 
\begin{equation}
\label{f}
P(x)\, =\, (\beta-1)\,x^{-\beta}\, ,
\end{equation}
where the minimum jump is $x=1$ and the jump 
can be to the left or to the right.

As before, the position of the new active site 
is obtained by jumping $x$ sites from the present one, i.e.\
the position of the new active site will 
be given by Eqs.\ (\ref{a2b},\ref{a2}), 
where now each copy of the system has its
own $x$.
Thus, the values of $x$ are un-correlated between the 
two copies of the system. The  new values of $f$
assigned to the active sites are the same. 
This choice results in a different behavior at 
the saturation regime. 

In  the  Random Walk limit, $\beta\gg 1$ in Eq.\ (\ref{f}), 
the distance Eq.\ (\ref{b22}), can be easily computed by
considering  Eq.\ (\ref{d}) together with the fact that 
$\sigma \sim t^{1/2}$ . This calculation yields 
\begin{equation}
\label{h}
\langle D(t)\rangle \,\sim\, t^{1/2}\, .
\end{equation}

In the general case  $\beta>1$, there is  still a power law
growth of the distance (\ref{b}) for intermediate times $1\ll t \ll N$. 
The exponent in Eq.\ (\ref{a1}) can be obtained  
using the fact that the mean-square distance $\sigma^2$ 
covered by a L\'{e}vy Walk behaves like 
\begin{eqnarray}
\label{i}
\sigma^2 (t) & \sim & \left\{
\begin{array}{lr}
t^2 & (1<\beta\le 2) \\
t^{4-\beta} & (2<\beta<3) \\
t\log t & (\beta=3) \\
t & (\beta>3)
\end{array} \right .  \, ,
\end{eqnarray}
in the long-time limit. As a consequence, one 
can also compute the so-called
dynamical exponent $z$ defined through $\tau\sim N^z$ where
$\tau$ is the time needed for the distance 
(actually the ratio  $\frac{\langle D(t)\rangle}{\langle D(1)\rangle}$) 
to reach the plateau. This time is
given by the time needed to cover all the lattice sites, 
if finite size effects
are not counted in. Comparison between Eqs.\ ({\ref{a1},\ref{d},\ref{i}) 
yields $z=\frac{1}{\alpha}$. 

In this simple model we have excluded any kind of 
correlation between the values
of $x$ extracted from Eq.\ (\ref{f}) and between  
the two replicas at $t=1$. Indeed, the dynamics is 
 given by a generalized random walk and therefore the power-law behavior of
the growth is not related to the statistical properties of the
system. This is in fact the idea behind our toy model: We use it as a ``black
box'', not knowing what happens inside, we are only able to 
observe  a L\'evy Walk behavior of the activity. 
Our model is, by conception, a trivial
system that has as only purpose that of showing what are the consequences,
 in the context of damage-spreading, of a 
power-law behavior of the activity like the one observed in the
Bak-Sneppen model. 
 
As we shall see in the next section, the non-trivial 
properties of the self organized critical systems 
are hidden in the value of the exponent $\alpha$ in
Eq.\ (\ref{a1}). Furthermore,  the averaging
 procedure leading from
Eq.\ (\ref{b}) to Eq.\ (\ref{d}) plays a very important role in these
non-trivial systems. 

Before moving onto the analysis of the BS model, let us 
discuss  in more detail which terms are contributing to the 
computation of $\langle D(t) \rangle$ via Eq.\ (\ref{d}). 
In the ring,  we have defined $\langle D(t) \rangle$
 by considering the behavior of $\sigma(t)$, which is a
physical quantity related only to the behavior of the activity in one single
system. In general, considering that the two replica
might be correlated, we need to update Eq.\ (\ref{d}) to 
\begin{equation}
\langle D(t)\rangle\, =\, \langle D(1)\rangle\, {\bar{n}}_{cov}(t)\, ,
\label{d1}
\end{equation}
where ${\bar{n}}_{cov}(t)$ is the average number of {\em different} 
sites covered by both system and copy. More precisely, suppose that 
at time $t$, 
the activity has covered $\sigma_1$ and $\sigma_2$ different sites 
in $R_1$
and $R_2$ respectively. Then, 
the function ${\bar{n}}_{cov}(t)$ is given by
\begin{equation}
{\bar{n}}_{cov}(t)\, =\, \langle 
\sigma_1+\sigma_2 - \sigma_{1,2} \rangle\, .
\label{d2}
\end{equation}
where $\sigma_{1,2}$ represents the number of sites 
covered in both systems (i.e.\ 
the {\it covering overlap} between system and copy). 
In the case of the ring, for large $N$ and $t\ll N$, 
the overlap on the rhs of Eq.\
(\ref{d2}) is empty ($\sigma_{1,2} \equiv 0$)
 and Eq.\ (\ref{d1}) reduces to Eq.\ (\ref{d}). In the case
of the Bak-Sneppen model instead, this intersection cannot be empty 
even in the thermodynamic limit. As a consequence, the exponents 
predicted from Eq.\ (\ref{i}) have to be considered as an upper 
bound for those observables in
systems with non-trivial correlations.
\section{The Bak-Sneppen Model}
\label{bsmodel}
As mentioned above, in its simplest version the BS model describes  an  
ecosystem as  a collection of $N$ species on a one dimensional
lattice. To each species corresponds a fitness described by a number $f$  
between $0$ and $1$. 
For simplicity, one considers  periodic boundary conditions. 
 The initial state of the system is defined by assigning to each site $j$ a 
random fitness $f_j$ chosen from a uniform distribution. 
The dynamics proceeds in three basic steps:
\begin{enumerate}
\item{} Find the site with the absolute minimum fitness 
on the lattice (the active site) and its two nearest neighbors.
\item{} Update the values of their fitnesses 
by assigning to  them new random numbers from a uniform distribution. 
\item{} Return to step 1.
\end{enumerate}
After an initial transient that  will be of no interest
to us here, a non-trivial critical state is
reached. This critical
state, characterized by its statistical 
properties, can be understood as the {\em fluctuating balance} between two
competing ``forces''. Indeed, while the 
random assignation of the values, together with the coupling, 
acts as an entropic disorder, 
the choice of the minimum acts as an ordering
force. As a result of this competition, at the stationary state
the majority of the $f_j$ have values
above a certain threshold $f_c=0.66702(1)$ \cite{BS93}. In other words, the
distribution function of the $f_j$'s can be asymptotically approximated by
\begin{equation}
\label{c2}
\eta_1 (f)\,  = \, \frac{1}{1-f_c}\, \Theta(f-f_c)\, ,
\end{equation} 
where $\Theta(f)$ is the Heaviside function.
Only a few will be below $f_c$, namely those 
 belonging to the
running avalanche (see \cite{BS93,PMB96} for a detailed discussion).
Proceeding by analogy with the previous cases, once the system is at
the critical state,
we produce two identical copies $B_1$ and $B_2$ 
and find the minimum (the active site).
Then, in $B_2$ we
swap the value of the minimum 
fitness with the fitness of some other site chosen at random
(note that if $N$ is big enough, the fitness in the other site will certainly
be above threshold).
After that, the evolution of the Hamming distance given by Eq.\ (\ref{b}) 
is studied. In the evolution of both system and 
copy the same random numbers are used. Here, the length of the jumps
in the position of the active site follows a power-law distribution given by
Eq.\ (\ref{f}) with $\beta\sim 3.23$ \cite{BS93}. 
At variance with the case of the ring
discussed  above, we cannot expect the behavior shown in
Eqs.\ (\ref{d}) and (\ref{i}). Indeed, in the BS model 
the jumps posses strong spatial correlations, with large 
probability  of returning to sites already covered in
previous time-steps. As a consequence, the behavior of the number of different
sites covered in one single system cannot be given by Eq.\ 
(\ref{i}) but leads to $\sigma(t)\sim t^{\mu}$ for $t\gg 1$ \cite{BS93},
with $\mu = 0.4114 \pm 0.0020$ \cite{Grassberger95}. 
Moreover, as we consider the two copies $B_1$
and $B_2$ we immediately realize that the two systems are also strongly
correlated and  consequently one obtains an even smaller
exponent
 leading to 
\begin{equation}
\label{d3}
\langle D(t)\rangle \, \sim\,  t^{\alpha=0.32}\, .
\end{equation}

As we mentioned at the end of Sec.\ \ref{general}, 
the behavior of Eq.\ (\ref{d3}) can be understood in the framework 
of Eq.\ (\ref{d1}). In fact, the decrease in the value of $\alpha$ is given 
 by the appearance  of $\sigma_{1,2} \neq 0$ in  Eq.\ (\ref{d1}).

Indeed, the fact that we are using the same sequence of random
numbers implies that, if the system is big enough the absolute
minimum in one system is the absolute minimum also in the copy. Therefore, if
the absolute minimum is among those sites which have not yet been covered by
the activity, the three terms in the rhs of Eq.\ (\ref{d2}) will have the same
behavior. If the minimum is instead one of the newer values put in the system
after perturbation, its position on the lattice may be different in the two
replica but the three functions in the rhs of Eq.\ (\ref{d2}) grow slowly or
even do not grow at all. This observation is confirmed by the irregular
behavior of $D(t)$ in just one single realization. In fact, it is the averaging
 procedure the one that produces  finally a 
smooth curve. It should be noted that, 
as it is clear from its definition, the behavior of the
intersection is strongly correlated to the behavior of the other two sets and
therefore the average in the rhs cannot be split into the sum of the
independent averages. 
As a consequence we should  expect a smaller exponent
with respect to the one
obtained for $\sigma(t)$.

The initial distance can be computed using Eq.\ (\ref{c11}). To do that we need
the distribution function of the value of the minimum. Extensive numerical
simulations indicate that this distribution can be approximated by 
\begin{equation}
\label{c1}
\eta_2 (f)\, = \, \left( 3-\frac{9}{2}\, f\right)\,\Theta(\frac{2}{3}-f)\, ,
\end{equation}
where the threshold has been put equal to $2/3$. 
Inserting (\ref{c1}) and (\ref{c2})  in Eq.\ (\ref{c11}) we  obtain
$\langle D(1)\rangle\sim \frac{11}{9 N}$. Since $\langle
D(t\to\infty)\rangle$ takes into account all sites on the same footing,
this saturation value can be 
obtained from Eq.\ (\ref{b1}) with $\rho_1=\rho_2=\eta_1$. The distribution 
$\eta_1$
comes from Eq.\ ({\ref{c2}), and the saturation value is $\langle
D(t\to\infty)\rangle\sim\frac{1}{9}$. Therefore, as in the case 
of the ring,
the saturation value does not depend on the size of the system while the
initial distance does. Thus, the normalized distance reaches a plateau that
must scale with $N$, as our numerical simulations show (see Fig.\ \ref{fig2}).

Coming back to the dynamical exponent $z$, we find that it still 
follows the prediction $z=1/\alpha$ 
as in the case of the ring. For the BS model one obtains, following the above
described prescription,   $z\sim 3.12$, instead of
$z\sim 1.6$ determined in \cite{TCT98}. This value $z\sim 3.12$ coincides
 reasonably well with the one  obtained from the collapse plot (Fig.\
\ref{fig3}). 
The reason for the discrepancy between our results and those 
presented in  \cite{TCT98} can be traced back to the effects
 of time-rescaling on the
(normalized) growth function. Indeed, let us assume we use 
 a different time-scale,  and 
consider the case in which we make a measure of $D(t)$
every $\nu$ time-steps (instead of every time-step), 
where $\nu$ is distributed according to a certain
function $P(\nu)$. The rescaled distance $\tilde{D}(t)$ will be given by
\begin{equation}
\label{eqti}
\tilde{D}(t)\, =\, \int\mbox{d}\nu\, P(\nu)\, D\left(
(t-1)\langle \nu\rangle\,+\nu\right)\, ,
\end{equation}
where $\langle \nu\rangle= \int\mbox{d}\nu\nu P(\nu)$ is the average number of
time-steps between two consecutive measurements. The choice made in
\cite{TCT98} corresponds to $P(\nu)=\delta(\nu-N)$. 
It is easy to see that,  in this case,  the growth 
exponent for $\tilde{D}(t)$ is still $\alpha$, 
 but the measured dynamical exponent is given by $1/\alpha -1$. 
In principle,  one could imagine more 
complicated distributions $P(\nu)$ for the
measuring time. In  particular, if $P(\nu)$ did not have a finite first
moment (as would be the case, for instance, 
if $P(\nu)$ corresponded to  the avalanche distribution)
Eq.\ (\ref{eqti}) would yield $\tilde{D}(2)\approx N$, i.e.\ the rescaled
distance would  saturate almost immediately \cite{comment3}.\\

In order to study the size dependence of $\alpha$ we have 
performed a set of measurements for different system sizes and then
extrapolated the value of $\alpha$ to infinite size. We get an asymptotic 
value $\alpha_{asym}=0.40(1)$ (see Fig.\ \ref{fig4}).

At this point, it is worth discussing what happens in higher dimensions.
The high-dimensional Bak-Sneppen model has been
extensively studied in \cite{DMV98}, 
where the behavior of the exponent $\mu$ for
the growth of the quantity $\sigma(t)$ has been computed until the mean-field
regime $\mu = 1$ has been reached. In the framework of the damage-spreading,
taking into account the correlations as discussed above, 
one  expects  
the exponent $\alpha$ to follow  a similar pattern, 
reaching the value $\alpha=1$ in the mean-field case. 
These mean-field results can also be
obtained in the random-neighbors case \cite{FSB93}. 
However, from the point of view of damage spreading, 
the random nearest neighbor case presents a complication. 
There is an ambiguity in the choice of the
neighbors. Indeed,   their absolute positions on the lattice
should be either the same in the two copies of the system
 or taken at random in an uncorrelated fashion. 
In both cases, each one of the two copies will behave
normally, but the behavior of the distance will  be completely
different. Indeed, in the first case the distance 
between the two systems will
never grow, while in the latter case the behavior 
of the distance resembles
that of the ring with uniformly distributed jumps.

\section{BS Model without noise}
\label{deterministic}
Through the use of different maps (chaotic as well as non-chaotic), 
it was shown  \cite{DVV97,DVV98}
that the random updating is not a necessary requirement to have
SOC. Moreover,  as long as the updating rule  is chaotic the
system does not change the universality class,  i.e.\ 
all the exponents are
the same as in the case of random updating.
This means that the system is able
to self-organize at a higher level: It takes 
into account the temporal correlation 
(or the average time spent in every site) by increasing
the threshold, so as to have the same statistical 
properties  \cite{DVV97,DVV98}. 
As a consequence, all equations and relations  
derived for the original BS model
are still valid for all the cases 
with chaotic updating.
The stationary distribution of the 
fitnesses, on the other hand,
follows a different pattern. Indeed, 
the position of the threshold as well as
the exact shape of the stationary 
distribution depends on the actual 
form of the updating rule.

To calculate the value of $\langle D(1)\rangle$ 
one needs to specify
the nature of the perturbation. Using the same methods as 
in the random updating case,
i.e.\ to 
swap the position of the
minimum in $B_2$ with any site taken at random,  
at $t=1$,
the average (initial) distance $\langle D(1)\rangle$ can be 
obtained from Eq.\ (\ref{b22}), 
where $\eta_i$ is the distribution function (at $t=1$)
for the variable $f^i\in B_i$.

When one considers the chaotic updating, Eqs.\ (\ref{b22},\ref{b1}) 
 are still valid. The only difference is that one needs to find
the stationary distributions that correspond to the actual map. 
For instance, for the tent map as well as for the Bernoulli map
the distribution coincides with that of the 
random updating (except for the
value of the threshold in the Bernoulli case) \cite{DVV97,DVV98}.
As an example, we substitute the random updating
with the logistic map
\begin{equation}
f_i(t+1)=b f_i(t)(1-f_i(t))\, ,
\label{logimap}
\end{equation}
where $i$ runs over the minimum and its nearest neighbors and $b$ 
is a parameter (that we set to the value $4$ which is at 
the threshold of the chaotic phase). Inserting the estimated distributions
obtained in \cite{DVV97,DVV98} one obtains as first approximation
$D_{asym}\sim 0.087$ and the same
exponent $\alpha$ as in the random updating.

In Fig.\ \ref{fig5} one can 
see the evolution of the distance for the case of a Bak-Sneppen model with
random and logistic updating, using the perturbation defined in 
\cite{TCT98} (flipping of the minimum fitness with another 
fitness chosen at random). 
Simulations with other
chaotic maps give the same exponents too: This 
reflects the observed fact that
chaotic maps do not change the universality class of the model
\cite{DVV97,DVV98}. 
In the inset we show the scaling of $ND_{asym}(N)$ versus $N$ 
for both random and logistic updating. A linear 
fit gives as estimation of the 
plateau $D_{asym}=0.112(5)$ for random updating and $D_{asym}=0.088(5)$ 
for logistic updating, in good agreement with our analytic estimate.

\section{parallel BS model}
\label{parallel}
The $d$-dimensional parallel version of the BS model (PBS)
\cite{SD96}, where parallel updating has been introduced,
has been found to have an exact mapping 
onto $2d$ Directed Percolation.
In fact, the avalanche time distribution in the PBS model is equivalent to the
cluster distribution in DP and the threshold for PBS is equivalent 
 to the critical probability in DP.

The dynamics in the PBS is different from the one in the (usual) Bak-Sneppen
model due to the introduction of  parallel updating:
\begin{enumerate}
\item{} Find the site with the absolute minimum fitness $f_{min}$
on the lattice (the active site) and its two nearest neighbors.
\item{} Update the values of their fitness 
by assigning to  them new random numbers 
from a uniform distribution. 
\item{} Search for all sites on the lattice with fitness $f < f_{min}$ and
update them together with their nearest
neighbors. Repeat the search until there are no sites left with $f < f_{min}$.
\item{} Return to step 1.
\end{enumerate}

The difference between the PBS model and the original BS model lies in
step 3. In the extremal version, once the minimum and its nearest neighbors
 are updated one looks
for the new minimum, and consequently the number of updated sites per time
step, $U_t$, is constant ($U_t \equiv 3$). In the parallel version, instead,  
this number will follow a complex temporal evolution during 
an avalanche with, in general, $U_t \geq 3$. In fact, the distribution 
of the number of updated sites per
time step (inside an avalanche) shows a nearly flat distribution, 
with a upper cutoff, whose value is comparable with 
the system size \cite{CVV99}. Due to this saturation, 
 one observes that $D(t)$ grows in time faster than in
the BS case and that finite size effects are also much stronger.

Since the disorder is stronger in the parallel version ($U_t \geq 3$), 
one expects the equilibrium point 
to be displaced towards the completely disordered value. In fact, 
for the PBS model, $f_c \approx 0.5371(1)$ \cite{SD96}. 
This results can be obtained both numerically  
and analytically by mapping the model onto
directed percolation.

To study the behavior under perturbations of the PBS model, we 
 follow the same procedure as for the extremal case.   
We produce two identical copies $B_1$ and $B_2$ of  the system of size $N$ in 
the critical state, and  find the minimum (the active site).
Then, we introduce a slight perturbation in  $B_2$
and  follow the evolution of the Hamming distance Eq.\ (\ref{b})
in time.
Since this quantity has strong fluctuations, 
we  consider  the average 
$\langle D(t)\rangle$ over different realizations of 
the initial values of the $f_j$. In particular all the simulations 
presented here, are the result  of 
averaging over $10^2$ realizations. The simulations are performed 
with both random and deterministic (logistic) updating rules.
Let us begin by discussing the results obtained for random updating. 
As mentioned above, $D(t)$ may depend on the internal
correlation of the system and on the correlations 
between the two copies. 
In the $1D$ BS model, due to the choice of unit of time, which allows only 
a number O(1) of sites to be updated, the growth rate must 
give an exponent $\alpha\! <\! 1$
and stop at a certain time $\tau\sim N^z$, 
at which a  crossover to a saturation 
regime appears. 

In the PBS case, too, $D(t)$ reaches, 
 after an initial power law growth (as in the BS case), a well defined  
plateau (see Fig.\ \ref{fig8}). 
However, due to  the faster parallel dynamics, the value 
of the exponent $\alpha$ is found to be larger than 
the one for the the extremal non-parallel case, 
as shown in Table \ref{tab1}. 
Moreover, there seem not to be 
dependence of the exponent $\alpha$ on the
system size $N$. Notice that our exponent 
differs from the one obtained in
\cite{HWD97} for the Domany-Kinzel model in the context of 
DP, onto which the PBS can be mapped. 
This discrepancy is due to the choice of time-scale for the measure 
of the distance. Indeed, in the Domany-Kinzel model case only one 
active (occupied) site per 
time step can be updated, together with its neighbor, 
thus resulting in a dilated 
time scale with respect to the study
presented here. Thus, in order to compare the two models one has to 
establish the relationship between the two time-scales. 
This is not easy to realize for the 
Hamming distance. In fact, according to this interpretation, 
 equal times for the two copies on the Monte-Carlo parallel 
time-scale are not equal times on the DP-like time-scale. To 
circumvent this problem, we realized a set of PBS simulations 
(with system size $N=2000$) for a single copy, computing at every 
Monte-Carlo (parallel) time step $\delta t=1$ the number $n_{act}(t)$ of 
sites below threshold (which is itself time-dependent). 
This defines the temporal increment 
for the DP-like time scale $\delta t_{DP}=\delta 
t \cdot n_{act}(t)=n_{act}(t)$. 
The effective 
DP time at the MC step $t$ is thus connected 
to the effective time at MC step $t-1$ by the relation:
\begin{equation}
t_{DP}(t)=t_{DP}(t-1)+n_{act}(t).
\label{tt}
\end{equation}
Then, we mediated over different realizations of the 
dynamics, obtaining the scaling of the effective DP time 
with the MC time of the simulation. The result, shown in 
Fig.\ \ref{fig8bis}, is that $t_{DP} \sim t^{\zeta}$ 
with $\zeta=1.41(2)$. From this, if we assume that $t_{DP}$ is
 the equivalent of the DP time-scale for PBS, we can combine the 
scaling law for $D(t)$ and that for $t_{DP}$ to get the 
effective scaling exponent $\alpha^{*}$ for the Hamming distance 
with respect to the DP time-scale $t_{DP}$:

\begin{equation}
\alpha^{*}=\frac{\alpha}{\zeta}=\frac{0.47}{1.41}=0.33(1).
\label{tDP}
\end{equation}

This value is quite near to the DP exponent 
$\alpha_{DP}=0.314(1)$ \cite{HWD97}. 

At any given time step $t$ during an avalanche, 
the average growth of $D(t)$ is connected to the mean number 
of sites $\sigma(t)$ covered by the activity in each system according to Eq.\
(\ref{d}). 
In Fig.\ \ref{fig10}, $\sigma(t)$ exhibits a power law behavior, 
$\sigma(t)\sim t^{\mu}$. The values of the  scaling exponent $\mu$ 
at different sizes $N$ are always bigger than the corresponding 
values of the exponents for the Hamming distance (Tab. \ref{tab1}). 
This is expected
since the correlations between the two copies in $D(t)$, if present, 
can only decrease the value of $\alpha$ with respect to $\mu$ 
\cite{VV98}. 
The asymptotic value of the Hamming distance shows quite a strong and 
persistent dependence on the system size $N$ and converges 
to an asymptotic value only logarithmically in $N$ (see inset 
of Fig.\ \ref{fig8}). Then, in order to get the real 
value of the plateau, one has to go to the thermodynamic 
limit $N=\infty$, once the plateau has been reached. This is 
realized by a logarithmic extrapolation of the data for 
different sizes. The value of this plateau can be 
obtained  in terms of the asymptotic stationary fitness 
distribution $\rho (f)$ from Eq.\ (\ref{b1})
where $\rho_i(f)$ is the normalized distribution 
function (at $t=\infty$) for the variables $f^i\in B_i$, 
at a given system size $N$. 
The dependence of the plateau $D_{asym}(N)$ on $N$ must then be 
related  to the shape of the stationary fitness distribution. 
The fitness  distribution has, in fact,  a  flat tail 
below the threshold $f_c$, 
which disappears only logarithmically in $N$ 
as the system size is increased. 
In the limit $N\to\infty$ the distribution $\rho_i(f)$ 
is given by Eq.\ (\ref{c2})
with $f_c=0.5371$. By substituting the distribution $\rho_i$ thus obtained
into Eq.\ (\ref{b1}) we get $D_{asym}(N=\infty)=D_{asym}=0.1543$ as 
estimation of the plateau in 
the thermodynamic limit.
Turning back our attention to Fig.\ \ref{fig8}, one can see that 
this result is 
consistent with the extrapolated numerical 
value $D_{asym}\sim 0.14(2)$. 

We have also performed  a similar  analysis for the PBS model 
with a deterministic updating, in which the new fitnesses are 
 obtained by iterating the logistic map, namely
\begin{equation}
f_i(t+1)=b f_i(t)(1-f_i(t)),
\label{logi}
\end{equation}
where $f_i(t)$ is the fitness of site $i$ at time $t$, 
and $b$ is a 
parameter set to the value $4$ \cite{DVV97,DVV98}. 
We observe the same qualitative 
behavior for all the quantities studied in 
the random updating case. The fitness distribution is 
however different since
it  is influenced by the invariant measure of the map, as
pointed out in \cite{DVV97,DVV98}. In the present case, 
the fitness distribution is
strongly picked around $f=0$ and around $f=1$, for finite sizes $N$. 
The distribution is of course
not symmetric and in the large $N$ limit, all the $f_j$ 
are above a threshold 
$f_c=0.55(1)$ \cite{Note3}. 
The value of the plateau $D_{asym}$ converges, in the limit 
$N\to\infty$, to 
$D_{asym}\sim 0.15(2)$. By inserting  
the fitness distributions thus obtained  in Eq.\ (\ref{b1}),
we obtain  an  analytic estimation,
$D_{asym}=0.1556$, that is indeed very close  to 
the random updating analytical value, and in agreement 
with our  numerical results. 
This is reasonable, since the threshold of the parallel 
BS with deterministic rule is very near the threshold of the 
random updating case. Although we performed, for the computation 
of the plateau,  simulations up to system size $N=8000$, the 
need of a logarithmic 
extrapolation towards $N=\infty$ prevents us from obtaining 
a precise numerical estimation of the plateau. The values of 
the exponents
$\alpha$ and $\mu$ show no substantial differences with respect 
to the
random updating case, as is the case for the extremal BS model 
\cite{CVV98},
thus confirming the robustness of $\alpha$ with respect to 
different updating rules.

\section{on the initial perturbations}
\label{pert}
One point remains, however, that needs to be 
studied. The definition of the
initial perturbation in the replica in \cite{TCT98} is too restrictive.
In fact, by considering as initial perturbation 
\begin{equation}
\tilde{f}_i=f_i + \epsilon g_i\, ,
\label{otherpert}
\end{equation}
where $\epsilon$ is a small positive number, we can take several choices for
$g_i$ 
without altering the exponent $\alpha$. We considered four different
implementations of $g_i$:
\begin{description}
\item[(a)] $g_i = \psi(t)$ where  $\psi(t)$ is a 
random number between 0 and 1.
\item[(b)] $g_i = \psi(t)$ where  $\psi(t)$ is 
a random number between -1/2 and
1/2.
\item[(c)] $g_i =\psi(t)$ for $i$ corresponding to the minimum and zero
otherwise. 
\item[(d)] $g_i=constant=1$ for all sites.
\end{description}
This kind of perturbation 
allows us to tune the initial mean 
distance $D(t)$, to any arbitrarily small
value  
depending on $\epsilon$ in Eq.\ (\ref{otherpert})
(contrary on the flipping introduced in \cite{TCT98}, 
which gives a fixed, size dependent, mean initial distance). 
In case (d), for example, (which is the case we will use in the
analysis below) the initial  distance 
is $D(1)=\epsilon$, independent on $N$,
and the plateau will be  
independent on $N$ too. This characteristic of the global perturbation  
is useful to explore  
some properties of the model with deterministic
updating and is more in line with the standard techniques 
of damage spreading problems.

We have performed the analysis with the 
prescription (\ref{otherpert}) also in
the case in which the model has a deterministic microscopic rule, like an
updating with the logistic map. 
In this case, the numbers in each replica will not be the same 
since the map will be applied on different numbers. In Fig.\ \ref{fig6}  we 
show the behavior of $D(t)$ for random and logistic updating, where we used 
perturbation (\ref{otherpert}) with the implementation (d). 
The exponent 
does not change: In both cases a value of $\alpha$ around the value $0.32$
found in \cite{TCT98} is obtained. The plateau of
$ND(t)$ (we computed $ND(t)$ to reduce 
statistical fluctuations) for the case
of global perturbation, scales linearly 
with the system size $N$, as shown in
the inset of Fig.\ \ref{fig8bis}, with a slope $a=D_{asym}=0.114(5)$ and
$a=D_{asym}=0.089(5)$ respectively for random updating and logistic
updating. The good agreement with the rough analytic estimate
obtained above and with numerical results for the flipping shows that 
the value of the plateau, too, does not depend on the kind 
of perturbation applied. 
Consequently, the distance $D(t)$ has a plateau $D_{asym}$
independent on the size, as it happens with the flipping. 
These results do not
change if we use the implementations $(a),(b),(c)$ 
of the perturbation (\ref{otherpert}).

The case of the logistic map with perturbation (\ref{otherpert}) is
particularly  
interesting from a different point of view. Since the map 
is chaotic, one could expect that on small scale distances ($D(t)\ll 1/N$) 
the chaoticity of the maps dominates and the distance of the two copies 
grows exponentially. For bigger distances 
instead, the dynamics is dominated 
by the damage spreading (and thus the critical properties of the BS model) 
and the distance grows as a power law. This is exactly what we find 
numerically. In Fig.\ \ref{fig7} we show
numerical simulations of the BS model with deterministic updating with 
the logistic map (with parameter value $4$) and random perturbation, for 
different system sizes and initial distances. 
The distance 
$D(t)$ has an initial 
exponentially growing phase with a Lyapunov exponent $\lambda=0.49(1)$. 
This Lyapunov exponent is smaller than the Lyapunov exponent 
of a single logistic map with $b=4$, which actually is $log 2=0.693147...$. 
Indeed, the microscopic dynamical rule of the system changes the values of 
the minimum fitness and its nearest neighbors, and the Lyapunov exponent 
we measure arises from the interaction 
between the three maps applied to the 
three different numbers.

\section{Conclusions}
\label{conclusions}
We have shown that the power-law behavior of 
the distance Eq.\ (\ref{a1}) has its origins in the 
behavior of the mean-squared distance covered
by the activity. The first consequence is that $\alpha\le 1$. The second is
that the internal correlations of the jumps, governed 
by Eq.\ (\ref{f}), and the
strong correlations between the two copies, can severely 
slow down the growth of
$\langle D(t)\rangle$,  thus leading to 
smaller exponents with respect to the
ones predicted in Eq.\ (\ref{i}). Since an analytic derivation of the
exponents characterizing the critical properties of the BS model is yet to be found,
we had to base our work on numerical results. Nevertheless, 
by  rewriting Eq.\ (\ref{a1})  into the more appropriate form given by Eqs.\
(\ref{d1}-\ref{d2}), we have
been able to study the mechanisms leading to it. 
In the framework of these equations the occurrence of the
plateau can be also easily understood and we have computed its value.

Our theory assumes 
that neither in the BS case nor in the case of
the ring will 
 the distribution of the variables $f$ change during the
measurement of $\langle D(t)\rangle$ (we exclude the transient).

Moreover,  we have studied the characteristics of damage spreading in a 
parallel version of the Bak-Sneppen model, and compared them to those of 
the extremal 
version. In general, the PBS model exhibits a faster 
evolution towards a stationary state, which is strongly dependent on the
system size $N$.  

Finally we have confirmed the picture presented in \cite{DVV98} according to which
the application of deterministic (chaotic) maps does not change the
universality class of the model. Furthermore,  we have shown here 
that the analysis attempted
in \cite{TCT98} and \cite{VV98} is independent of the
microscopic details of the system,  and of the prescription of the initial
perturbation. This of particular importance if one looks at these
results within  the  framework of standard damage-spreading. The possibility to
apply this analysis to such different versions of the Bak-Sneppen model, together with 
the good agreement with the results for damage-spreading in Directed
Percolation, gives us confidence that these techniques can be extended to other
Self-Organized Critical models.

\begin{table}
\begin{centering}
\begin{tabular}{|c|c|c|c|c|c|} \hline
$N$ & $500$ & $750$ & $2000$ & $4000$ & $8000$\\ \hline
$\alpha$ ran. up. & $0.48(1)$ & $0.48(1)$ & 
$0.47(1)$ & $0.49(1)$ & $0.48(1)$ \\ \hline
$\alpha$ log. up.  & $0.47(1)$  
&$0.47(1)$  &$0.49(1)$ & $0.48(1)$ & $0.48(1)$\\ \hline
$\mu$ ran. up. & $0.63(1)$ & $0.63(1)$ & 
$0.65(1)$ & $0.65(1)$ & $0.65(1)$\\ \hline
$\mu$ log. up.  & $0.63(1)$  
&$0.64(1)$  &$0.63(1)$ & $0.65(1)$ & $0.65(1)$\\ \hline
\end{tabular}
\caption{Values of the exponents $\alpha$ and $\mu$ for different 
sizes and updating rules.}
\label{tab1}
\end{centering}
\end{table}

\begin{figure}[h]
\centerline{\psfig{file=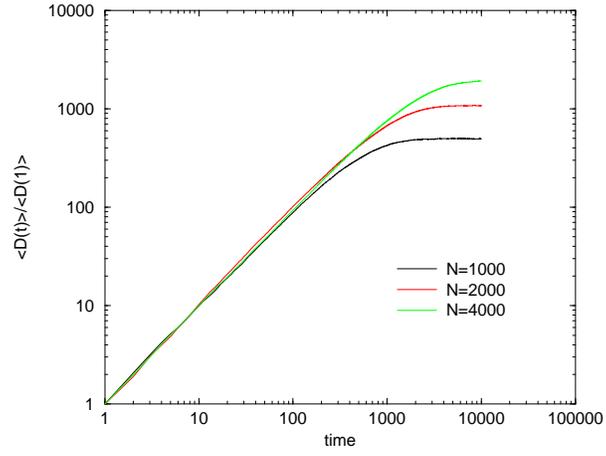,height=7cm}}
\caption{Distance $D(t)$ between the two replicas for three different sizes
 $N$ for the uncorrelated ring model.}  \label{fig1}
\end{figure}

\begin{figure}[h]
\centerline{\psfig{file=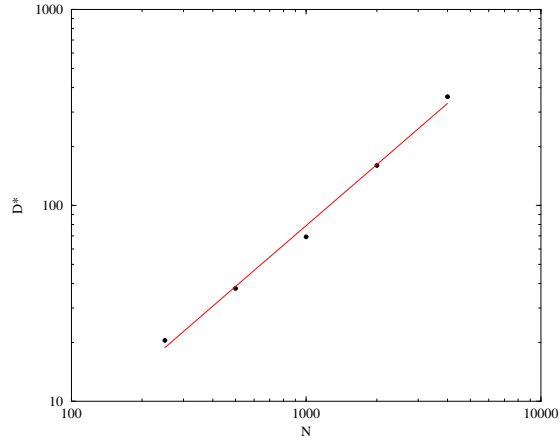,height=7cm,angle=-90}}
\caption{Scaling of the long time plateau 
$D^* = \frac{\langle D(\infty)\rangle}{\langle D(1)\rangle}$ 
as a function of the number of sites $N$ for the BS model. The best fit 
yields an exponent of $1.03(4)$.} 
 \label{fig2}
\end{figure}

\begin{figure}[h]
\centerline{\psfig{file=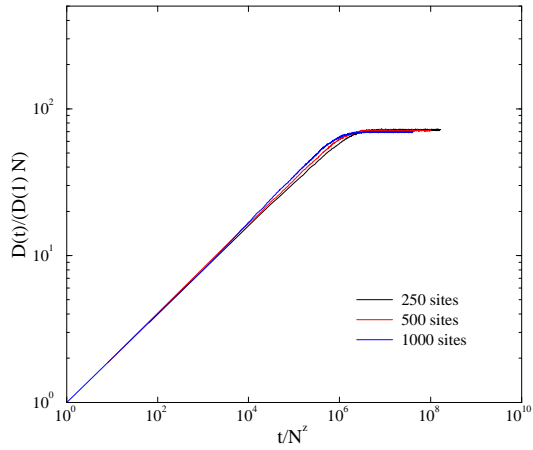,height=6cm}}
\caption{Collapse plot of the data for the evolution of the distance, in
 the BS case. The best fit yields $\alpha=0.32(3)$ and $z=3.0(2)$.} 
 \label{fig3}
\end{figure}

\begin{figure}[h]
\centerline{\psfig{file=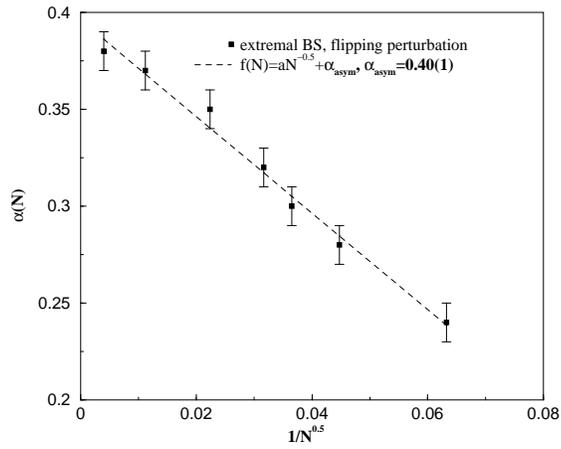,height=6.0cm}}
\caption{Extrapolation of $\alpha$ for the BS model with random updating.} 
\label{fig4}
\end{figure}

\begin{figure}[h]
\centerline{\psfig{file=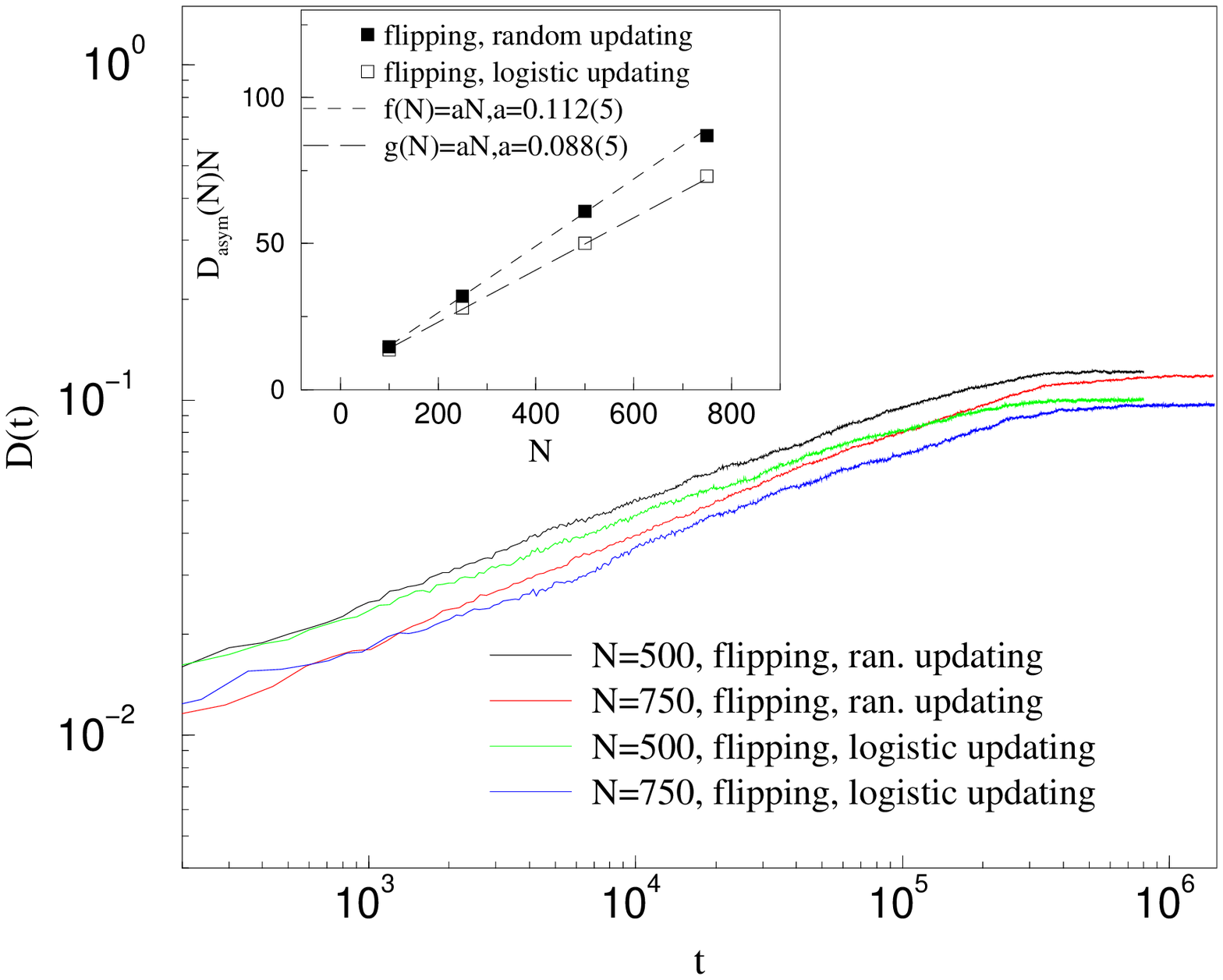,height=6cm}}
\caption{$\log_{10}$-$\log_{10}$ plot of $D(t)$ for the BS model 
with flipping, with random updating and deterministic updating 
with logistic map, for different system sizes. The value of the 
plateau actually depends on the updating rule, but it does not 
depend on the system size. Inset: A plot of the 
plateau of $ND(t)$ versus $N$ fits very well with a linear 
scaling, showing that the plateau of the Hamming distance $D(t)$ 
is size independent.}  
\label{fig5}
\end{figure}

\newpage

\begin{figure}
\centerline{\psfig{file=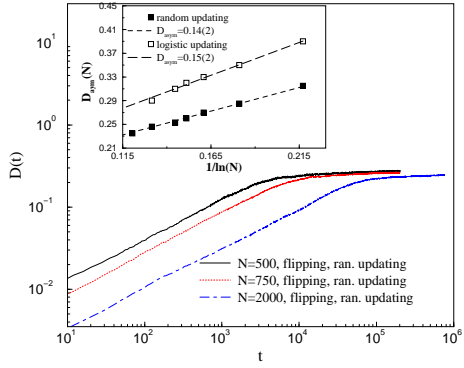,height=5cm}}
\caption{The $log_{10}-log_{10}$ plot of the Hamming distance $D(t)$ in the 
parallel BS model, versus the time $t$. 
Inset: logarithmic extrapolation of 
the plateau as a function of $N$, for both random and logistic 
updating rules. The infinite size 
value $D_{asym}$ obtained is in agreement with our analytical estimate.}  
\label{fig8}
\end{figure}

\begin{figure}[h]
\centerline{\psfig{file=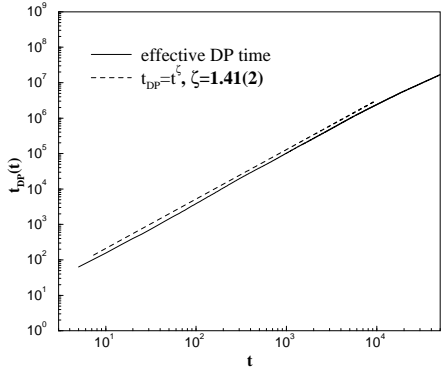,angle=0,height=5cm}}
\caption{Scaling of the effective DP-like time $t_{DP}$ with the 
Monte-Carlo time $t$, on a $log_{10}-log_{10}$ scale. We get a power 
law behavior with exponent $\zeta=1.41(2)$.}  
\label{fig8bis}
\end{figure}

\begin{figure}
\centerline{\psfig{file=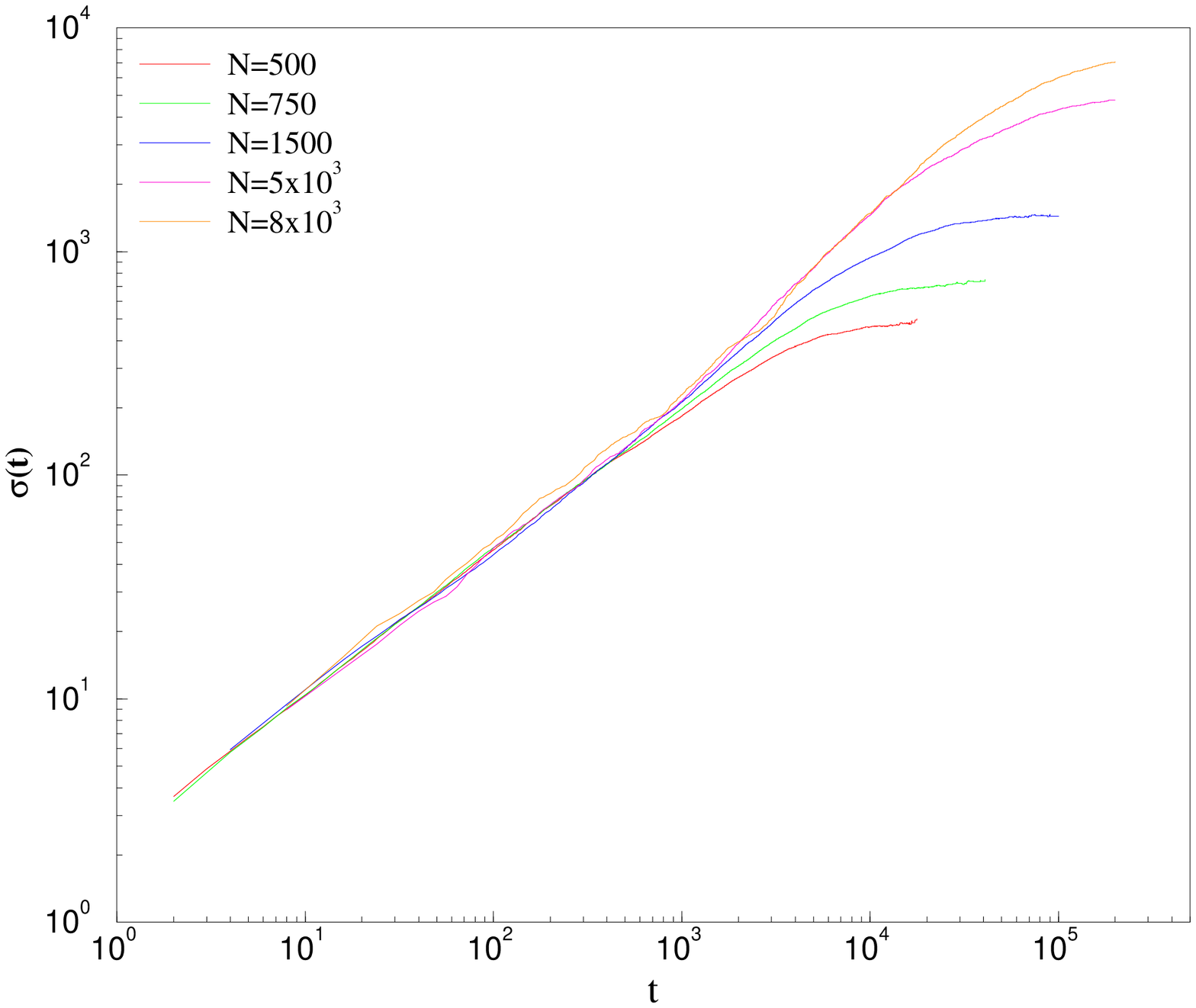,angle=-90,height=5.5cm}}
\caption{$log_{10}-log_{10}$ plot of the number $\sigma(t)$ of sites 
covered during the evolution of a single copy of the system, 
versus the time $t$, for random updating rule.}  
\label{fig10}
\end{figure}

\begin{figure}[h]
\centerline{\psfig{file=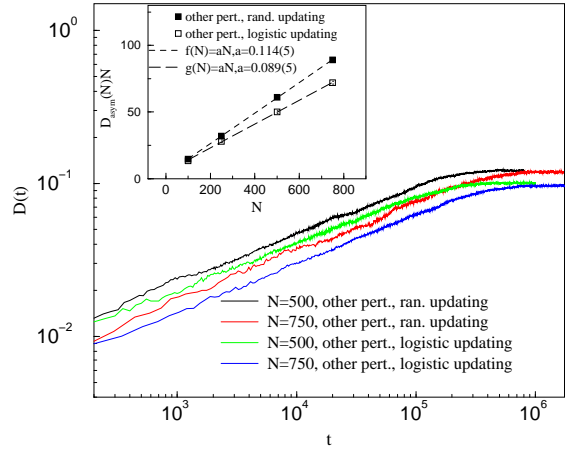,height=7cm}}
\caption{$\log_{10}$-$\log_{10}$ plot of $D(t)$ for the 
BS model with the perturbation (\ref{otherpert}), 
implementation (d), with random updating and deterministic 
updating with logistic map, for different system sizes. 
The value of the plateau actually depends on the 
updating rule, as in the flipping case. 
Inset: A plot of the plateau of $ND(t)$ 
versus $N$ gives the same findings of the case 
of flipping.}  
\label{fig6}
\end{figure}

\newpage

\begin{figure}[h]
\centerline{\psfig{file=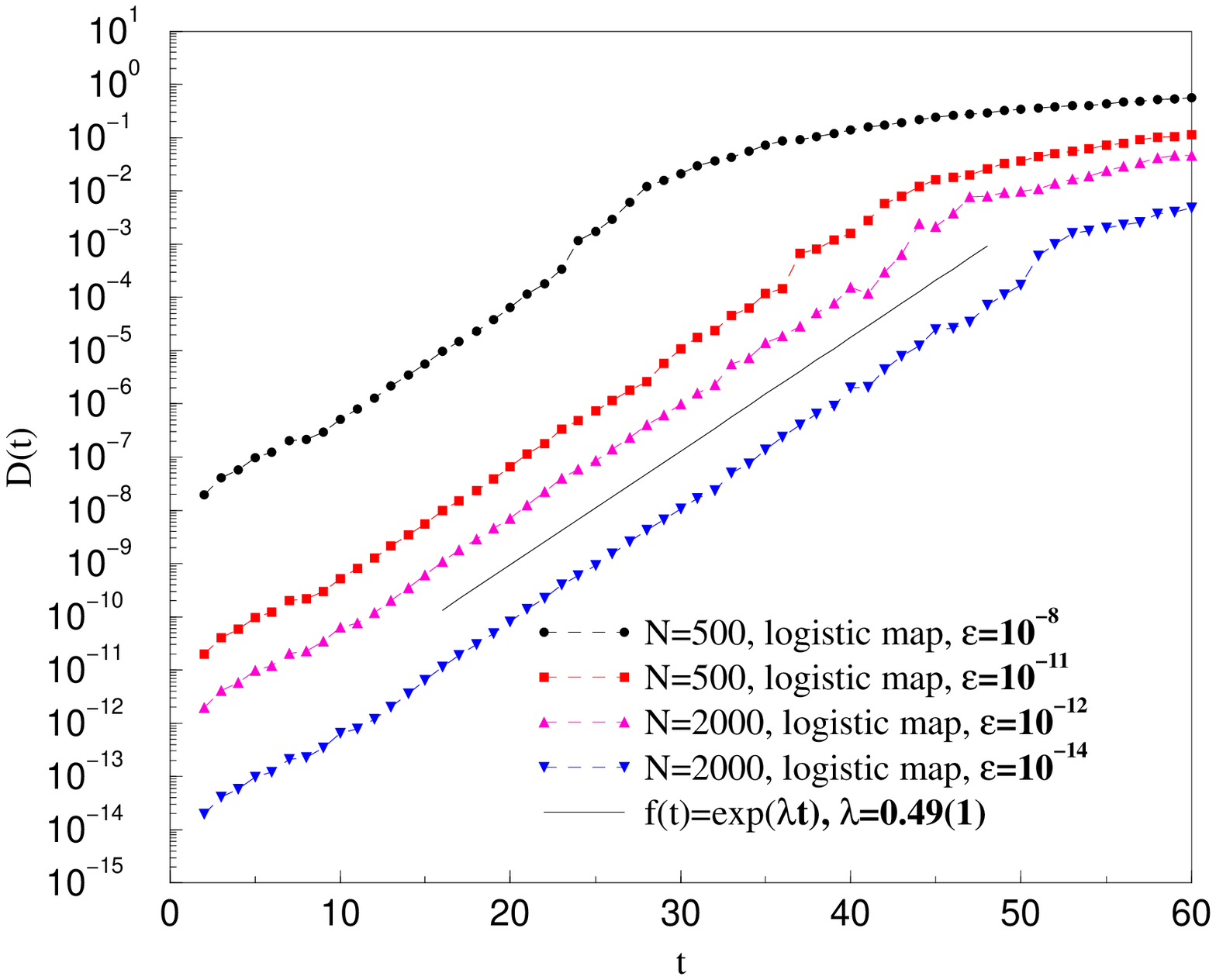,height=7cm}}
\caption{$linear$-$\log_{10}$ plot of $D(t)$ for the BS model with 
perturbation (\ref{otherpert}), implementation (d), and 
updating with a logistic map with parameter $b=4$, for 
different system sizes and 
mean initial distances $\epsilon$.}  \label{fig7}
\end{figure}


\begin{thebibliography}{99}
\bibitem{comment3} Since, 
for finite $N$ every distribution
has  a cut-off, and therefore every moment is finite, this line of
reasoning applies only in the limit $N\rightarrow \infty$. A more detailed
analysis of this point has been presented elsewhere \cite{CVV98}.
\end{thebibliography}
\end{document}